\documentclass[conference]{IEEEtran}
\IEEEoverridecommandlockouts
\usepackage{cite}
\usepackage{amsmath,amssymb,amsfonts}
\usepackage{algorithmic}
\usepackage{graphicx}
\usepackage{textcomp}
\usepackage{xcolor}
\usepackage{cite}
\usepackage{amsmath,amssymb,amsfonts}
\usepackage{algorithmic}
\usepackage{graphicx}
\usepackage{textcomp}
\usepackage{xcolor}

\usepackage{cite}
\usepackage{amsmath,amssymb,amsfonts}
\usepackage{algorithmic}
\usepackage{graphicx}
\usepackage{textcomp}
\usepackage{psfrag} 
\usepackage{cuted}
\usepackage{float}
\usepackage{graphicx} 
\usepackage{array}
\usepackage{bbm}
\usepackage{mathrsfs}
\usepackage{epstopdf}
\DeclareMathAlphabet{\mathpzc}{OT1}{pzc}{m}{it}
\usepackage{graphicx}
\usepackage{amsmath}
\usepackage{caption}
\usepackage{subcaption}

\def\BibTeX{{\rm B\kern-.05em{\sc i\kern-.025em b}\kern-.08em
    T\kern-.1667em\lower.7ex\hbox{E}\kern-.125emX}}
\begin{document}

\title{Data-driven Thermal Modeling for Electrically Excited Synchronous Motors - A Supervised Machine Learning Approach\\
}

\author{\IEEEauthorblockN{Farzaneh Tatari, Davis Trapp, Jason Schneider, Mohsen Mirza Aligoudarzi}
\IEEEauthorblockA{\textit{Control Systems, Drive System Design Inc.} \\
Farmington Hills, MI, USA \\
farzaneh.tatari@drivesystemdesign.com; davis.trapp@drivesystemdesign.com;\\ jason.schneider@drivesystemdesign.com; mohsen.aligoudarzi@drivesystemdesign.com;}
}

\maketitle

\begin{abstract}
This paper proposes a data-driven supervised machine learning (ML) for online thermal modeling of electrically excited synchronous motors (EESMs). EESMs are desired for EVs due to their high performance, efficiency, and durability at a relatively low cost. Therefore, obtaining precise EESM temperature estimations are significantly important, because online accurate temperature estimation can lead to EESM performance improvement and guaranteeing its safety and reliability. In this study, in addition to the default inputs' data, EESM losses data is leveraged to improve the performance of the proposed ML approach for thermal modeling. Exponentially weighted moving averages and standard deviations of the inputs are also incorporated in the learning process to consider the memory effect for modeling a dynamical thermal model. Using the experimental data of an EESM prototype, the performance of ordinary least squares (OLS) method is evaluated through a complete training, testing and cross-validation process. Finally, simulation results will provide the key performance metrics of OLS for EESM thermal modeling.
\end{abstract}


\section{Introduction}
Due to the trend towards sustainable mobility and electric vehicles (EVs), thermal analysis of motors has recently received great attention. Electric motors must operate within a safe temperature range to protect their components from excessive heat while maximizing their produced power and torque. Therefore, it is highly beneficial to know the motor temperature in real time to optimize the motor's performance and prevent the motor overheating and failing. Temperature sensor installment can provide accurate information about the motor's components thermal state in real time. However, it can be difficult and expensive to install sensors in some critical parts, such as the rotating rotor parts. Therefore, electric motor temperature estimation through thermal modeling is extensively studied \cite{wallscheid2021thermal} to provide a more feasible and cost-effective solution.

Accurate thermal modeling leads to a reliable and cost-effective temperature estimation which can be employed online in the temperature monitoring of electrically excited synchronous motors (EESMs) to enhance the safety, control performance, and effective cooling of the EESM.
Moreover, having a precise estimation of motor temperature, the control module can perform better at critical operating points, especially when power derating is necessary \cite{wallscheid2017derating}.
Thermal methods such as computational fluid dynamics and  high heat equation finite element analysis are highly accurate, but due to their complex and expensive computations, are not appropriate for real time temperature estimations \cite{wallscheid2021thermal}. 

Generally, there are three main methods for electric motor temperature estimation: indirect methods, lumped-parameter thermal networks (LPTNs), also known as direct methods, and supervised machine learning.
Indirect methods track electrical motor parameters that are sensitive to temperature. For example, the resistance of the motor's windings increases as the motor heats up. By inferring the resistance of the windings, it is possible to estimate the temperature of the motor.
However, these methods suffer from increased machine losses and sensitivity to machine parameters that lead to inaccurate modeling and severe estimation errors \cite{wallscheid2014real}. On the other hand, LPTNs (direct methods) are a type of thermal model that represents the motor as a network of electrical nodes and can be used to directly estimate the temperature. Each node represents a different part of the motor, such as the stator, rotor, or bearings. The nodes are connected by resistors, which represent the thermal resistance between the different parts of the motor. 
LPTN methods which are categorized to white-box and gray-box methods, simplify the partial differential heat transfer equation into a set of ordinary differential equations. 
White-box LPTN is based on motor design and material information, and does not utilize empirical data for parameter tuning and Gray-box LPTN integrates the general LPTN framework and empirical data-based model identification. LPTN parameters are difficult to calculate accurately using only motor design and material information and recorded data from motor prototypes can help to estimate more accurate parameters.

Supervised machine learning (ML) which is the interested method in this paper, is a type of artificial intelligence, used to learn the relationship between the inputs and outputs of a system. ML can be employed for motor thermal modeling where the input variables could include the motor's speed, torque, current and etc. and the output variable would be the motor's components temperatures. Supervised ML techniques go further than LPTNs and do not rely on physical assumptions. This makes ML models more generalizable to different motor types and application scenarios.
ML models are often called black-box approaches because they can be difficult to interpret. However, it is possible and advisable to incorporate some expert domain knowledge into the model design. This can help to improve the accuracy of the model and make it more interpretable.
Data-driven ML temperature estimation is a young and promising field and there exist several reasons for pursuing ML temperature estimation rather than traditional thermal modeling. The first is \emph{generalization}; where these models can be generalized to different motor types and application scenarios, while LPTNs are more specific to a particular motor type and operating condition.
Second is  \emph{automation}; where ML models can be automated to a high degree, since they do not rely on physical assumptions. This can save time and effort in the modeling process.
Third is  \emph{bypassing uncertainty}; because electrical motor temperature estimation models always come with some degrees of uncertainty, due to the inaccurate or unmodeled physical effects. However, ML thermal models project the available motor input data to the targeted output motor temperatures without requiring the complex physical relationships. This can help to reduce the uncertainty in the temperature estimation.

Some methods of ML for temperature estimation include linear regression, ordinary least squares (OLS), and multi-layer perceptron (MLP) \cite{wallscheid2021thermal}. In this paper, we aim to present an accurate data-driven ML thermal model for EESMs by employing OLS method.

EESMs are desired for EVs because of their high efficiency and power density, wide speed range and ruggedness. EESMs are very efficient, converting a high percentage of electrical energy into mechanical energy that allows EVs to travel further on a single charge. Furthermore, One big advantage of EESM is that they do not rely on expensive rare-earth magnetic materials since the rotor is wound with copper wire instead.

High power density of EESMs enables them to produce a lot of power for their size and weight and allows them to have a smaller and lighter motor, which can free up space for other components, such as batteries.
Moreover, EESMs' wide speed range and ruggedness enables them to operate at a variety of speeds without losing efficiency
and be well-suited for EV applications' demand. 
Furthermore, EESMs promise specific benefits for EVs such as good transient response (they can quickly respond to changes in torque and speed), low noise, and cost. 
Low noise contributes to a more comfortable driving experience and low cost of EESMs' manufacturing makes EVs more affordable.
As a result, EESMs are expected to play an increasingly important role in the future of EVs and their accurate thermal modeling can improve their performance and efficiency even further.

Most of the existing research for motor's thermal modeling focuses on permanent magnet synchronous motor (PMSM) or induction motors and there exists a few studies on EESM thermal modeling. The authors in \cite{wang2022low} presented a low-order lumped parameter thermal network for online thermal monitoring of EESMs motor where thermal resistances and capacities were identified separately. However, even data-driven low order LPTN models require extensive motor domain expertise and time. The work of \cite{spielmann2022transient} presented an EESM transient thermal lumped parameter model with forced air cooling for shape optimization. 
To the best of our knowledge, there is no study on data-driven ML thermal modeling of EESM. The lack of data-driven thermal modeling for EESM, which can significantly improve the performance of EESM in EVs, formed the key motivation of our research.  

The work of \cite{kirchgassner2021data} studied the accuracy of several ML methods for PMSM thermal modeling where OLS and MLP models outperformed other methods in terms of accuracy. Convolutional neural networks (CNN) were studied in \cite{kirchgassner2020estimating} for PMSM thermal modeling which reported strong results near to the precision of the
LPTN \cite{wallscheid2015global}. However, CNNs, required over 42 days to tune the internal weights
\cite{kirchgassner2020estimating} and had an inference time significantly higher than an
OLS model \cite{kirchgassner2021data}. Thermal neural network \cite{kirchgassner2023thermal}, resulted in the best performance compared to other methods by leveraging neural networks
to integrate the concepts of LPTNs and universal differential equations \cite{kirchgassner2023thermal}. However, since no data is reported, the thermal neural network is expected to have similarly high inference times and take remarkable effort to tune.
In this paper, we aim to present an accurate thermal modeling for EESMs using supervised ML for the first time, where we incorporate the nonlinearities of the data and calculated power losses to propose a powerful data-driven thermal modeling architecture for EESMs. 
OLS is the ML method that is employed for the EESM data-driven thermal modeling. 
Empirical data of a prototype EESM is leveraged to present an accurate thermal model.


\section{ML algorithm for EESM thermal modeling}

In this paper, ML algorithm, OLS, is considered and extended with calculated power loss data, which have a proven suitability for these types of applications \cite{kirchgassner2021data}. 
In every ML model, upon several multi-dimensional input observations $\boldsymbol{X}=\left(\boldsymbol{x}_1, \boldsymbol{x}_2, \ldots, \boldsymbol{x}_n\right)^{\top} \in \mathbb{R}^{n \times p}$ an associated one-to-one mapping to a target output variable $y \in \mathbb{R}^n$
is learned, where $p$ denotes the number of input variables and
$n$ is the amount of observations. 
In our data $X$, each column represents a single measurement from a sensor in the vehicle's drive train where the sensors are independent. Each row combines these measurements, representing a single instance of data. Because we can access this data through the electric control unit, no extra hardware needs to be deployed for machine learning models. The dependent variables are the EESM stator and rotor temperatures.

Inputs and target outputs for EESM thermal modeling in this research is listed in Table 1. 

\begin{table}[ht]
\caption{Thermal Model Inputs and Outputs}
\centering
\begin{tabular}{c c c c}
\hline\hline
 Parameter type &  Parameter name &  Symbol    \\ [0.5ex] 
\hline
{Inputs} &  &   \\ [0.5ex] 
&Ambient temperature&  $\vartheta_a$ \\
&WEG-In temperature& $\vartheta_{wi}$ \\
&WEG-Out temperature& $\vartheta_{wo}$ \\
&Oil temperature& $\vartheta_o$ \\
&Oil flow rate& $r_o$ \\
&Actual d-axis current&  $i_d$ \\
&Actual q-axis current & $i_q$  \\
&Actual f-axis current & $i_f$  \\ 
&Actual d-axis voltage & $u_d$  \\ 
&Actual q-axis voltage & $u_q$  \\ 
&Actual f-axis voltage & $u_f$  \\ 
&Motor speed & $n_m$  \\ 
&Motor torque & $\tau$  \\ [1ex]
\hline
Derived inputs &  &  \\ [0.5ex] 
&Current magnitude $\sqrt{i_d^2+i_q^2}$& $i_s$ \\
&Voltage magnitude $\sqrt{u_d^2+u_q^2}$& $u_s$ \\
&Apparent power $u_s \times i_s$& $S_e$  \\
&Motor speed and current interaction $n_m \times i_s$ & $ni$ \\
&Motor speed and power interaction $n_m \times S_e$& $nS$  \\ 
\hline
  Target Outputs & &  \\ [0.5ex] 
&Rotor windings temperature& $\vartheta_r$ \\
&Stator windings temperature& $\vartheta_s$  \\
\hline
\hline
\end{tabular}
\label{table:nonlin}
\end{table}

\subsection{Ordinary least squares}	
OLS models work by finding a linear relationship between the input data (features) and the desired outputs (targets). In our case, the objective is to predict the temperature at $q$ different points inside the electric machine where these predicted temperatures over time are denoted by $\hat{\boldsymbol{y}}$. This predicted temperature $\hat{\boldsymbol{y}}$ is obtained using another matrix,  $X$, which contains the known input features. The linear model between the input data and the desired outputs is represented as follows
\begin{align}
\hat{\boldsymbol{y}}=\boldsymbol{X} \hat{\boldsymbol{\beta}},
\end{align}
where $\hat{\boldsymbol{\beta}}$ is the unknown parameters' matrix. The following sum of squares cost function is minimized to find the optimal unknown parameters $\hat{\boldsymbol{\beta}}$,
\begin{align}
    \boldsymbol{J}(\hat{\boldsymbol{\beta}})=(\hat{\boldsymbol{y}}-\boldsymbol{X} \hat{\boldsymbol{\beta}})^{\mathbf{T}}(\hat{\boldsymbol{y}}-\boldsymbol{X} \hat{\boldsymbol{\beta}}). \label{cost}
\end{align}
Differentiating \eqref{cost} with respect to $\hat{\boldsymbol{\beta}}$ and solving the obtained equation leads to the optimal solution as follows,
\begin{align}    \hat{\boldsymbol{\beta}}=\left(\boldsymbol{X}^{\mathbf{T}} \boldsymbol{X}\right)^{-1} \boldsymbol{X}^{\mathbf{T}} \hat{\boldsymbol{y}}.
\end{align}
Unlike some other methods, OLS models don't require a step-by-step optimization process to find the optimal parameters. This makes them really fast to set up and run. Refer to \cite{hastie2009elements} for a more detailed explanation of OLS method.

\section{Feature engineering}
\subsection{Default inputs and targets}
Accurately predicting temperatures inside an electric machine relies on using inputs that are consistent across different machines. 
To address this challenge, we use only the features listed at the first section of Table 1 as inputs for our models. While this initial data provides valuable information, creating additional characteristics from this data (through a process called feature engineering) can significantly improve the model's accuracy. 
The middle section of Table 1 listed engineered features as they improved model accuracy throughout this study where similar engineered features have been employed in \cite{hughes2023real,kirchgassner2021data} for PMSM thermal modeling.  Finally, the targets are shown at the bottom section of Table 1.

\subsection{Machine losses}

Electric machines have three types of losses that can be calculated including copper losses, iron losses, and mechanical losses.
To consider these losses in the model, the simplest form of every loss equation is considered by removing all constant variables, as they do not impact the scaled inputs for the ML models. 
This simplified loss equations provide usable input features for the ML model.
\begin{enumerate}
    \item Copper losses (Winding losses): 
    These losses are caused by the resistance of the windings in the stator and rotor. Copper losses are proportional to the square of the current flowing through the windings, 
$$
\begin{aligned}
P_{\text {cu,s }} & =1.5 \cdot k_{\mathrm{r}}  R_s \cdot\left(i_{\mathrm{d}}{ }^2+i_{\mathrm{q}}{ }^2\right), \\
P_{\text {cu,r }} & =R_r \cdot i_{\mathrm{f}}{ }^2,
\end{aligned}
$$
where $i_{\mathrm{d}}, i_{\mathrm{q}}$ denote the d-part and q-part of the stator currents in the flux-oriented coordinate; $k_{\mathrm{r}}$ is the field coefficient representing the ratio of phase resistances in alternating and direct currents; $R_s$ is the phase resistance of the stator winding; $i_{\mathrm{f}}$ is the rotor excitation current; $R_r$ is the winding resistance in the rotor. 

 The copper losses formula can be simplified to the inputs
\begin{align}
    P_{\mathrm{cu,s}} \propto \left(i_{\mathrm{d}}{ }^2+i_{\mathrm{q}}{ }^2\right), \quad P_{\mathrm{cu,r}} \propto i_{\mathrm{f}}^2, \label{cop}
\end{align}
as $k_{\mathrm{r}}$, $R_s$, $R_r$ are assumed to be constant.

   \item  Iron losses (Core losses): 
Core losses are caused by the hysteresis and eddy current effects in the magnetic core of the motor.
Iron losses $P_{fe}$ occur in the core of electric machines due to the constantly changing magnetic fields. These losses can be broken down into two main types: hysteresis losses $P_h$ and eddy current losses $P_e$, where the following equation
\begin{align}
    P_{\mathrm{fe}}= P_h + P_e = k_{\mathrm{h}} f B_{\mathrm{m}}^2+k_{\mathrm{e}} f^2 B_{\mathrm{m}}^2, \label{Pfe}
\end{align}
is employed due to its simplicity and applicability. The coefficients $k_h$ and $k_e$ are assumed to be constant in \eqref{Pfe}. Therefore, knowing that frequency $f$ is
proportional to motor speed
\begin{align}
    f=\frac{\omega_{\mathrm{m}}}{2 \pi},
\end{align}
and 
\begin{align}
    n_{\mathrm{m}}=60 \frac{\omega_{\mathrm{m}}}{2 \pi},
\end{align}
the iron losses inputs for the model can be obtained as
\begin{align}
    P_{\mathrm{h}} \propto n_{\mathrm{m}} ,
\end{align}
and 
\begin{align}
    P_{\mathrm{e}} \propto n_{\mathrm{m}}^2 . \label{eddy}
\end{align}
Since calculating the
maximum flux density $B_{\mathrm{m}}$ is nontrivial, it assumed that $B_{\mathrm{m}}$ is constant. 

\item  Mechanical losses: These losses are caused by friction in the bearings and inside the motor. 
 As bearing losses $P_b$ is calculated with
\begin{align}
    P_{\mathrm{b}}=2 \pi n_m T_{fric},
\end{align}
the equation can be simplified down to just the rotational speed of the
rotor where $T_{fric}$ is assumed to be constant. Since
\begin{align}
    P_{\mathrm{b}} \propto n_{\mathrm{m}},
\end{align}
already exists in our inputs, this does not lead to a new feature.

Given the previous derivations, the additional loss features are copper losses in \eqref{cop}, and eddy current loss in \eqref{eddy}. 
The new loss features that are proportional to the machine losses are summarized in Table 2 .

\begin{table}[ht]
\caption{Derived loss features equations}
\centering
\begin{tabular}{llll}
\hline Loss Type & & Loss Feature Equation  \\
\hline Copper &  & $P_{\mathrm{cu,s}} \propto \left(i_{\mathrm{d}}{ }^2+i_{\mathrm{q}}{ }^2\right), \quad P_{\mathrm{cu,r}} \propto i_{\mathrm{f}}^2$  \\
Iron & Eddy & $P_{\mathrm{e}} \propto n_{\mathrm{m}}^2 $  \\
\hline
\end{tabular}
\end{table}
\end{enumerate}


\subsection{Moving averages}

While OLS models are useful, they can't inherently capture the dynamics of a model where things change over time. This can be important for modeling thermal systems, which constantly fluctuates.
However, by incorporating a memory of past conditions into the model's inputs, one can improve OLS ability to handle dynamic systems. This memory is created through a technique called feature engineering where not only the current data from Table 1 is fed to the model but also data from previous moments are also incorporated as inputs. This allows the model to learn how past conditions influence future temperatures, making it more effective for dynamic systems.

To implement this, the authors in \cite{kirchgassner2021data} and \cite{wallscheid2014real} utilised exponentially weighted moving averages (EWMA) and exponentially weighted moving standard deviations (EWMS) to improve the accuracy of static models. 

 To account for changing temperatures over time, the model considers two special calculations for each input value ($X$) at every time step ($t$). These calculations are EWMA ($\mu_t$) and EWMS ($\sigma_t$) computed as follows
\begin{align}
  \mu_t=\frac{\sum_{i=0}^t w_i x_{t-i}}{\sum_{i=0}^t w_i} ,\quad \sigma_t=\frac{\sum_{i=0}^t w_i\left(x_i-\mu_t\right)^2}{\sum_{i=0}^t w_i},  
\end{align}
where $w_i=(1-\alpha)^i$ with $\alpha=2 /(s+1)$ and $s$ being the span. 
Different span values can be employed to create several versions of the original data. Each version acts like a filter, highlighting different frequency components of the original signal.

In this study $s$ values representing 1, 5, and 10 min are chosen.
Incorporating these moving averages to every single input, the number of features is multiplied by a
factor of six. It is worth to note that moving averages were applied
to the data of every duty cycle individually. 
For all variables an initialization buffer is used to provide values for the moving averages for the samples before the first 1, 5, and 10 min time spans elapse, where the EWMA and EWMS of every variable in this interval are respectively set to the first recorded value and zero.  


\section{Experimental data and evaluation of the proposed method}
Our experimental data comes from a 3-phase, 190 kW EESM containing 54-slots/6-poles and operating at 10 Hz. 
The mentioned EESM test bench is shown in Fig. 1. 
\begin{figure}
	\centering
	\includegraphics[width=3.5 in,height=2 in]{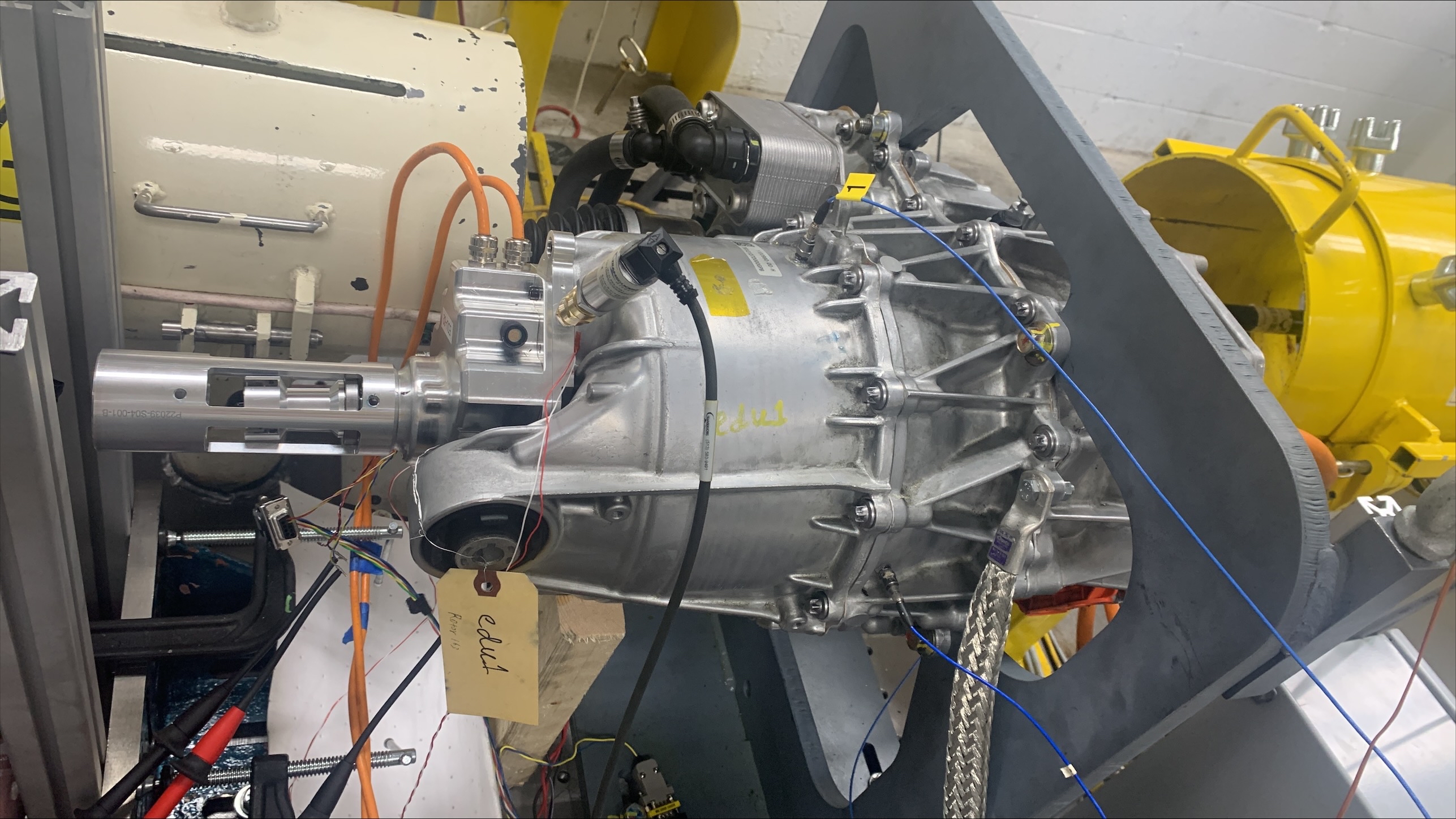}
	\caption{EESM in the test bench. The slip ring transferring the rotor temperature signal can be seen in the left of the image.}
	\label{figurelabel}
  \end{figure}
The employed dataset includes about 7 hours of recordings containing over 240 thousand multi-dimensional samples.
To capture crucial details for supervised learning, we placed sensors for measuring stator and rotor temperatures, where the position of these sensors on the EESM rotor and stator are shown in Fig.2.  
 \begin{figure}
	\centering
	\includegraphics[width=3.5 in,height=2 in]{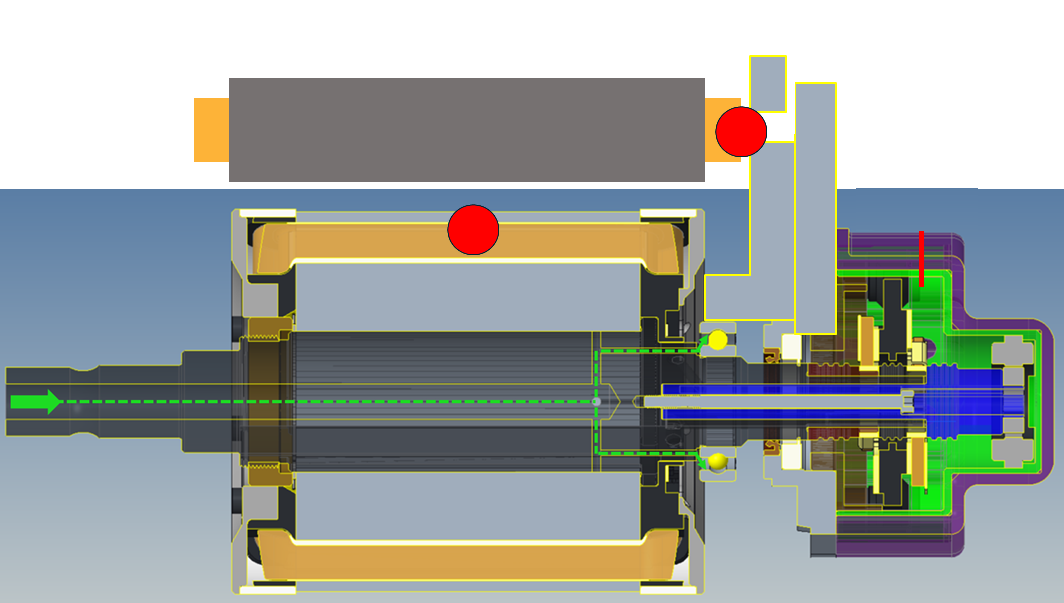}
	\caption{EESM rotor and stator temperature sensors' positions depicted with red circles.}
	\label{figurelabel}
  \end{figure}
Fig. 3 illustrates the torque-speed values of three measurement cycles of the experimental data set. As shown in Fig. 3, the entire collected experimental data contains samples where torque is within $[-150,200]$ (Nm) and speed is also limited to $[-1000,8000]$ (rpm) interval.
 \begin{figure}
	\centering
	\includegraphics[width=3.5 in,height=2 in]{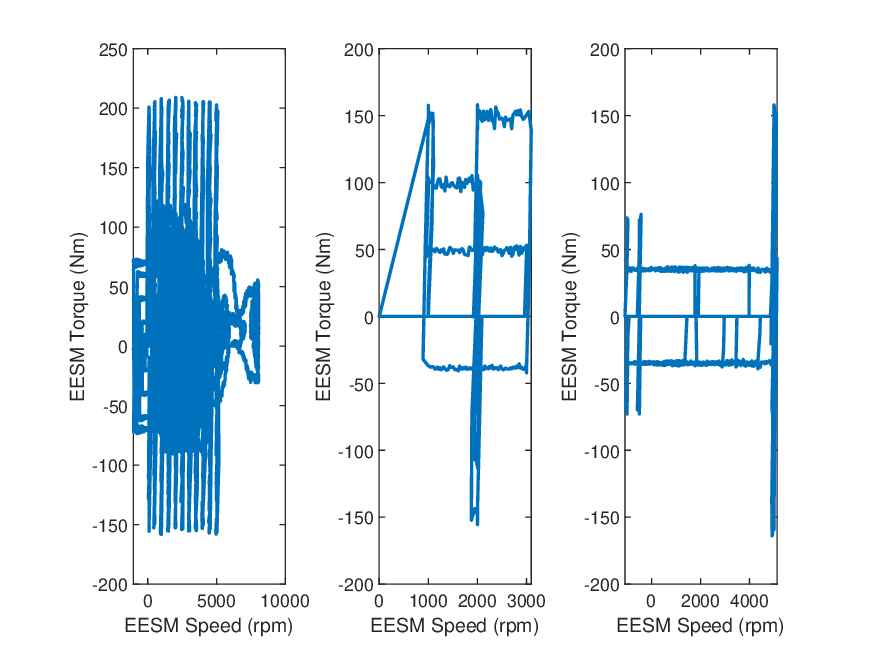}
	\caption{Toque-speed values of three measurement cycles. In total these sets make up 276 min of 400 min of experimental data.}
	\label{figurelabel}
  \end{figure}

\subsection{Training and testing pipelines}
Supervised machine learning relies on data-driven model building. After training, a separate validation step is crucial. Training involves optimizing model parameters through iterative calculations based on the training data. Validation, through a separate dataset, assesses how well the trained model generalizes to unseen data, preventing overfitting and assuring predictive accuracy on new examples.
$k$-fold cross-validation which is employed in this paper, is a way to test how well a machine learning model will work on new data. It does this by splitting the data into $k$ parts and training the model on $k - 1$ parts. The model is then tested on the remaining part of the data. This process is repeated $k$ times, with each part of the data used as the test set once. The average of the model's performance on the k test sets is then used as an estimate of its overall performance. Fig 4. depicts the $k$-fold cross-validation process.
 \begin{figure}
	\centering
	\includegraphics[width=3.2 in,height=1.5 in]{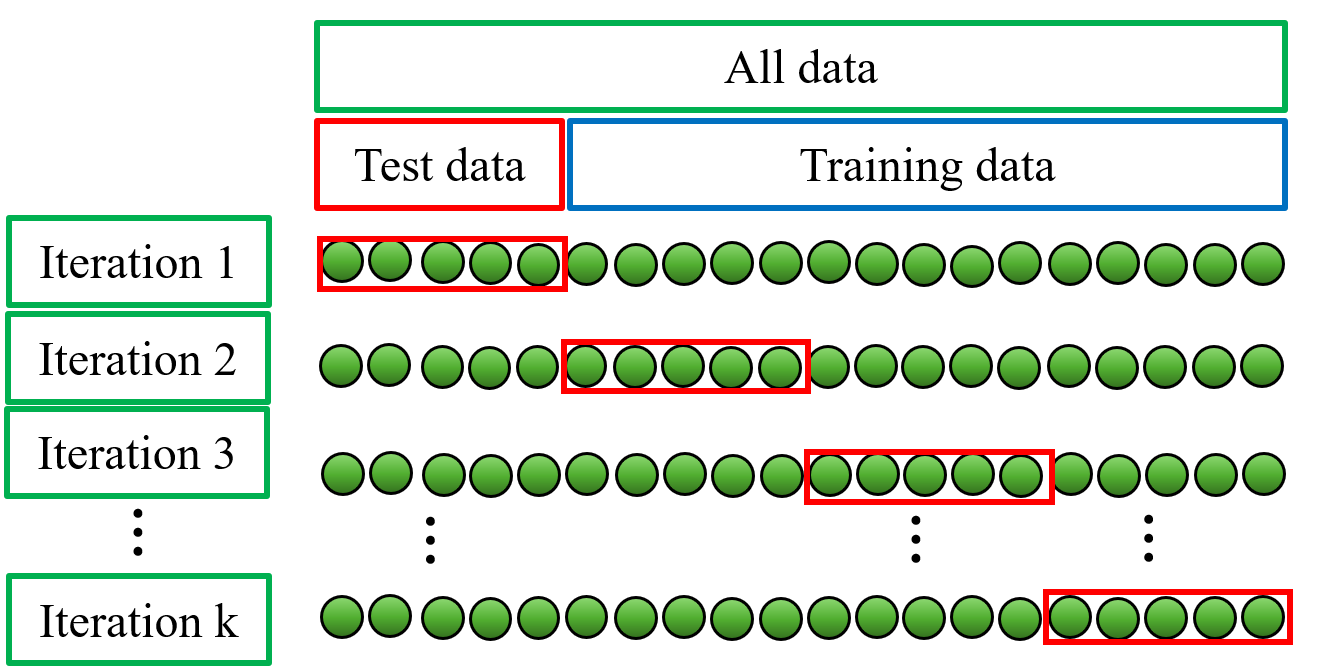}
	\caption{$k$-fold cross-validation.}
	\label{figurelabel}
  \end{figure}

Scores of the employed ML methods are reported using the mean squared error (MSE), mean absolute error (MAE), maximum absolute error (MaxAE)  between predicted sequence and ground truth. 
The test sets are used to compare the accuracy of the presented OLS model.

\section {Incorporating loss inputs results}
In this section, we will discuss the benefit of leveraging EESM machine loss data in its data-driven thermal modeling.

To compare the benefits of each loss derived input from Section 3, the
average score of the 10-fold cross validation is
compared.
The effects of each loss type are shown for the OLS model in Fig. 5 in terms of MSE average.

The incorporation of losses provides an improvement over the OLS thermal model. This might be because the additional nonlinear loss features could capture some complex relationships that the OLS without loss inputs could not fully handle. For OLS models, the best performance for both rotor and stator temperature estimation is found by incorporating all loss inputs which contain copper and iron losses which improves the MSE average for rotor and stator temperature estimation, respectively from 3.1013 and 0.5005 to 2.8983 and 0.4170. As shown in Fig. 5, incorporating copper and iron losses inputs in the OLS model individually has also improved the accuracy of the OLS for both rotor and stator temperature estimation. 
 \begin{figure}
	\centering
	\includegraphics[width=3 in,height=1.5 in]{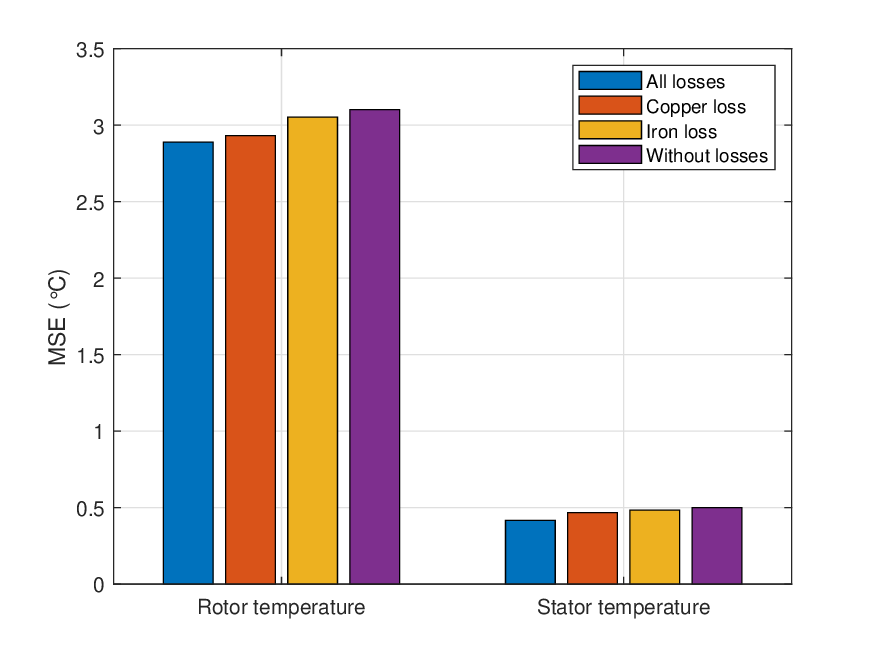}
	\caption{MSE of OLS over 10-fold cross validation results incorporating different loss inputs.}
	\label{figurelabel}
  \end{figure}
Following the results shown in Fig. 5, the loss included OLS model will
include all losses inputs containing copper and iron losses for both rotor and stator temperature estimations.
These results are also given in Table 3, where the complete OLS model
requires a maximum of 123 input features. The
number of input features is calculated as six times the number of
input features mentioned in the first section of Table 1 and three times both the number of derived inputs in the middle section of Table 1, and the number of loss features incorporated. It is worth to mention that every input (mentioned in the first section of Table 1) has two moving averages of EWMA and EWMS, each with three spans. Moreover, the derived inputs (mentioned in the middle section of Table 1) and losses inputs have one moving average, EWMA, each with three spans.  
\begin{table}
	\begin{center}
		\caption{MSE average comparison for 10-fold cross-validation and the number of features for different OLS EESM thermal models}
		
		\label{tab:table1}
		\centering
		\begin{tabular}{|m{1.5cm}| m{2cm} | m{1.9cm} | m{1.5cm} | } 
			\hline
			\text{OLS with} & \centering\text{Rotor Temp MSE}  & \centering\text{Stator Temp MSE} & \text{no. of features}\\
                \hline
			\text{All losses} & \centering\text{2.8983} & \centering\text{0.4170}& \text{123}  \\
			\hline
			\text{Copper loss} & \centering\text{2.9317} & \centering\text{0.4670}& \text{119}  \\
			\hline
			\text{Iron loss} & \centering\text{3.0526} & \centering\text{0.4842}& \text{115}  \\
			\hline
			\text{No loss} & \centering\text{3.1013} & \centering\text{0.5005}& \text{111}  \\ 
			\hline
		\end{tabular}
	\end{center}
\end{table}

\section {Simulation results}
In this section, the results of key performance metrics of enhanced OLS with loss inputs for EESM thermal modeling is presented where all data processing and model development is performed using MATLAB. 

Employing the experimental data and a 10-fold cross validation process, the MSE, MAE and MaxAE of the introduced enhanced OLS with loss inputs are given in Table 4. The EESM rotor temperature estimation using the enhanced OLS method has resulted in, the average MSE 2.8983, MAE 1.2797 and MaxAE 16.7639, and the stator temperature estimation using the same method has resulted in average MSE 0.4664, MAE 0.4693 and MaxAE 14.2433. The reported results show that for EESM thermal modeling, the enhanced OLS model had a good performance where the OLS model performs better for stator temperature estimation rather than rotor temperature estimation.   

Moreover, Fig. 6 shows the enhanced OLS model performance for estimating the EESM rotor and stator temperatures for a measurement cycle where this cycle was unseen during training process.

\begin{table}
	\begin{center}
		\caption{$k$-fold cross-validation testing results of OLS for EESM rotor and stator thermal modeling}
		
		\label{tab:table1}
		\centering
		\begin{tabular}{|m{0.8cm}| m{0.8cm}  m{0.8cm}  m{0.8cm} |m{0.8cm}  m{0.8cm}  m{0.8cm} | } 
			\hline
			\text{} & \multicolumn{3}{c|}{OLS rotor thermal model} & \multicolumn{3}{c|}{OLS stator thermal model}\\
                \hline
			\text{Iteration} & \centering\text{MSE} & \centering\text{MAE}& \centering\text{MaxAE} & \centering\text{MSE} & \centering\text{MAE}& \text{MaxAE}  \\
			\hline
			\centering {1} & \centering{2.7966} & \centering{1.2731}& \centering{16.4804} & \centering{0.4369} & \centering{0.4667}&  {6.7475}  \\
			\hline
			\centering {2} & \centering{2.9485} & \centering{1.2832}& \centering{17.1300} & \centering{0.4861} & \centering{0.4694}&  {16.5784}   \\
			\hline
			\centering {3} & \centering{2.8551} & \centering{1.2651}& \centering{17.4081} & \centering{0.4594} & \centering{0.4679}&  {16.0219}  \\ 
			\hline
			\centering {4} & \centering{2.9797} & \centering{1.2882}& \centering{16.7487} & \centering{0.4820} & \centering{0.4733}&  {16.5359}  \\ 
			\hline
                \centering {5} & \centering{2.9219} & \centering{1.2785}& \centering{16.1624} & \centering{0.4810} & \centering{0.4724}&  {16.0214}  \\ 
			\hline
                \centering {6} & \centering{2.8577} & \centering{1.2711}& \centering{16.6866} & \centering{0.4611} & \centering{0.4668}&  {15.9143} \\ 
			\hline
                \centering {7} & \centering{2.8725} & \centering{1.2762}& \centering{16.5210} & \centering{0.4719} & \centering{0.4718}&  {15.8191}  \\ 
			\hline
                \centering {8} & \centering{2.9926} & \centering{1.3018}& \centering{16.5075} & \centering{0.4736} & \centering{0.4686}&  {15.8014}  \\ 
			\hline
                \centering {9} & \centering{2.9295} & \centering{1.2831}& \centering{17.4231} & \centering{0.4573} & \centering{0.4689}&  {6.9296}  \\ 
			\hline
                \centering {10} & \centering{2.8290} & \centering{1.2763}& \centering{16.5712} & \centering{0.4551} & \centering{0.4674}&  {16.0639}  \\ 
			\hline
               \centering {\textbf{Average}} & \centering{\textbf{2.8983}} & \centering{\textbf{1.2797}}& {\textbf{16.7639}} & \centering{\textbf{0.4664}} & \centering{\textbf{0.4693}}&  {\textbf{14.2433}} \\ 
			\hline
		\end{tabular}
	\end{center}
\end{table}
\begin{figure}
	\centering
	\includegraphics[width=3 in,height=3. in]{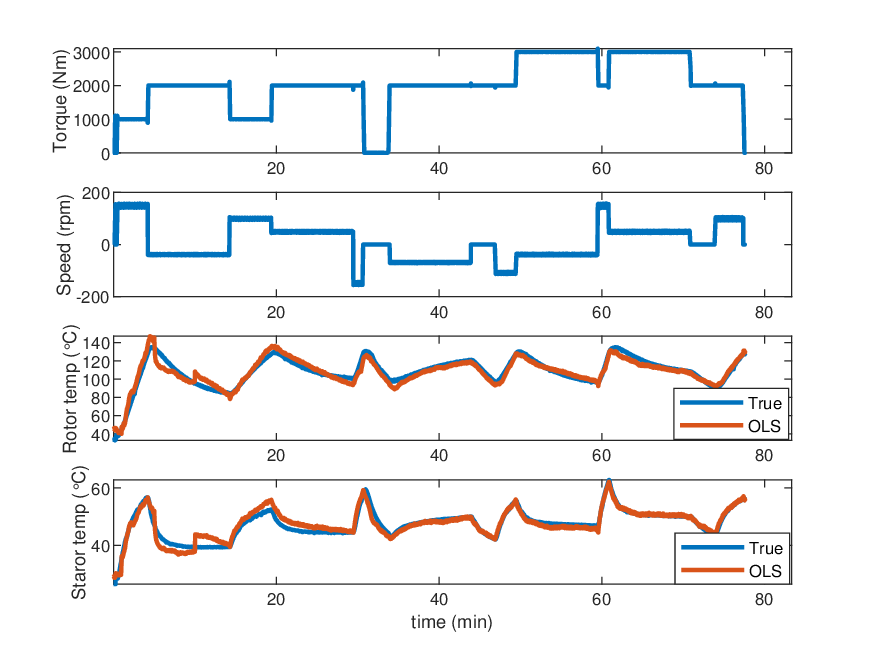}
	\caption{Motor speed, torque, rotor and stator temperatures estimations for a measurement cycle (test set).}
	\label{figurelabel}
  \end{figure}
  
\section{Conclusion}
This paper developed an ordinary least squares (OLS) method for online thermal modeling of electrically excited synchronous motors (EESMs) which can be emplyed to improve EESM performance, safety and reliability. Incorporating EESM losses inputs was also studied in this paper which improved the performance of the OLS thermal model. The dynamics of the EESM thermal model was captured with the OLS including exponentially weighted moving averages and standard deviations of the employed inputs. The training and testing of OLS was performed using experimental data where the final results of a 10-fold cross validation showed the performance of the given OLS for EESM thermal modeling.   

\bibliographystyle{IEEEtran}
\bibliography{IEEEabrv,ref.bib}

\vfill

\end{document}